# Optimizing Near Field Computation in the MLFMA Algorithm with Data Redundancy and Performance Modeling on a Single GPU

**Abdolreza Torabi1, Morteza Sadeghi2**
1- Associate Professor
2- Student of Algorithms and Calculations
*Sadeghi.morteza@ut.ac.ir*

**Abstract**
**The Multilevel Fast Multipole Algorithm (MLFMA) has known applications in scientific modeling in the fields of telecommunications, physics, mechanics, and chemistry. Accelerating calculation of far-field using GPUs and GPU clusters for large-scale problems has been studied for more than a decade. The acceleration of the Near Field Computation (P2P operator) however was less of a concern because it does not face the challenges of distributed processing which does far field.**
**This article proposes a modification of the P2P algorithm and uses performance models to determine its optimality criteria. By modeling the speedup, we found that making threads independence by creating redundancy in the data makes the algorithm for lower dense (higher frequency) problems nearly 13 times faster than non-redundant mode.**

## 1. Introduction

The FMM[1] and MLFMA [2] algorithms accelerate matrix-vector multiplication (MVM) in some scientific modeling problems in the form Aλ=B (λ is unknown). MLFMA reduces the complexity of MVM to O(N) in special cases [3]. It splits λ matrix to $\lambda_{Near}$which is called near-field and is calculated directly using kernel kernel and $\lambda_{Far}$which is called far-field and is approximated by hierarchicaly decomcoordination of kernel function. This hierarchical decomcoordination generates a implicit regular tree with multipoles as nodes. Each multipole represents the total field enters to or exits from a single mesh that contains several samples on studying surface. Solution of far field requires one pre-order traversing followed by an post-order traversing. Each transition between nodes or points requires specific type of operator. There are five types of operators named P2M, M2M, M2L, L2L and L2P for far-field calculation and only one operator for near-field named P2P.

This algorithm is popular due to its kernel independent and error controlable nature. However, when solving very large problems in heterogeneous clusters, problems arise in successive stages of computation. That is, calculations at each level of the tree wait for the higher or lower level to complete. This limitation introduces another problem related to the complex interactions between the multipoles placed on far processing nodes which makes network latency. Another problem is that the upper levels of the implicit tree are underutilized because there are few multipoles and they have to wait until other computational steps are completed. Pioneering works [8][9][10][11] focused on the key problem .i,e. efficiently distributing computations over cluster nodes which was not possible unless modifications on kernel computation. By shifting to frequency mode for higher level multipoles in implicit tree, network communications dropped significantly and utilization improved. There are many techniques like [11] that reduced computational redundancies and used GPU facilities to speedup calculation of M2L. Although their work is beyond the scope of this article, it gives an idea that modifying the algorithm of a particular processor can lead to further speed-ups of MLFMA.

Acceleration of P2P operators has not received much attention because it can be decomposed into smaller independent subtasks and processed simultaneously with far-field processing using more free resources (CPU). As mentioned in [7], the P2P operator is the second longest operator in his MLFMA.That is, it takes about 30% of the execution time.

Several works [4][5][6] attempt to reduce the complexity of running this operator on GPUs. Study [4] used shared memory and on-the-fly technique to reduces memory consumption and its access time thus became able to increas processing volume. For the P2P operator, they loads data on the thread scope registers which has low access time and low capacity. Its low capacity allows each thread to handle interactions of only 320 pairs of points. Using the Quadro FX5800 GPU and the Xeon E5520 CPU, speed of calculating near-field for problems between 2E7~7E7 points by this method is between 20 and 37 times faster than the CPU.

In [5] they focues on improving coalesced access to memory. They performed the interactions between points within the frame and points outside the frame separately. That is, for each point the receiver only checks its interactions with the points inside the box, and once its interactions with points outside the box. The second set of interactions produces a sparce matrix. Indices of this matrix are initially generated in the CPU and then used in the GPU. This separation eliminates the need for threads belonging to one block to access memory points associated with other blocks. They applied their method an an aircraft with 1E7 unknowns. The speed of P2P on four NVIDIA Tesla C2050 GPUs which were grouped using OpenMP-CUDA tool, is about 100 times faster than four Intel Xeon W3550 CPUs.

In [6] NGIM algorithm, which is similar to MLFMA, is implemented using the same technique of seperating coalesced and non-coalsced access of P2P operator. In addition, they used on-the-fly technique and computed the neighboring indices at runtime ,i.e, they increase runtime computations to reduce memory volume created by precalculation of indices and variables. The algorithm is evaluated on the GPU GTX480 and Xeon X5248 CPU .Speedup of P2P in this method is related to density of samples inside meshes. For boxes with 16 points inside them is between 200 and 260 times faster than the CPU while for boxes with 32 points, it is about 380 to 470 times faster, and 500 to 640 times faster for boxes with 64 points.

The work [4] conducted experiments on the fourth problems with different tree sizes (number of points) and levels, but did not indicate the effect of point density on the speed or technique used. The work [6] also listed the results of nine experiments with different tree levels and different density points. Their results shows in a problem with a fixed tree level, point density has a direct effect on the speed of execution. But they did not expand its effect in detailes. These prior works choise for work volume shared among threads was experimentally or intuitivly and not with precise calculations. One major reason is GPU speed is high enough so that analytical evaluation of work distribution seems not mandatory.

## 2. Our Contribution

As discussed in the previous section, improving the GPU memory access model has been a challenge for P2P operators. This issue arises from shared memory items that are accessed by multiple device threads. For this problem, we propose to redundant the data and make the processing threads completely independent of each other. This introduces additional overhead in data collection phase in CPU and increases volume of transferred data between GPU and RAM ,i.e, any restructuring of data in GPU affects the way data is collected in CPU. To measure how the presented solution affects the overall speedup, we use analytical models. It means that we predict the overall speed obtained from data restructuring based on the algorithm parameters used in the data restructuring algorithm.
While these models approximately predict speed, they guide us to find the optimal parameter space where P2P with redundant restructured data is still faster than non-redundant case. Our prediction models show that the distribution of sample points across boxes is the most important parameter for acceleration of our technique. The importance of sample density relative to boxes has also been mentioned in previous works [6]. Based on this finding, we performed additional tests on density and found that the presented redundancy technique is favorable for high-frequency problems or problems with a small number of samples per box.
The kernel selected for this work is electrical potential function on a two-dimensional PEC where the source and target (radiating and receiving) points are randomly distributed on the surface.
Our contribution brifely is listed below:

- Solving weak coalesced access to GPU memory by injecting redundancy and making thread's data seperated from each other ,i.e, restructuring GPU's input data
- Approximating the total speedup of presented technique using analythical models
- Modeling cache miss rate by presenting a novel metric for measuring locality of memory access requests
- Use analythical models to find the optimal parameter space in which presented solution works efficiently
- Presenting a simple improvement for P2P algorithm that is easy to model and analyse

In next part of this article, the simple implementation is presented by its analythical modeling, then the same is done for improved version. In the fourth section, accuracy of models is examined using empirical data. Then based on models, the efficiency of presented technique is tested on problems with various densities while running on a single GeForce 1050. Finally in section 5,6 conclutions and future works are discussed briefly.

## 3. Presented Technique

Since there is not a unique global implementation of P2P kernel, a simple implementation which is easy to implement and model is selected as a reference implementation. Then a small improvement is made on it and the effect of this improvement on speedup is predicted. The first implementation is called indexing method and the second implementation is called repetition method on this article. To simplify analytical models, none of conventional techniques such as overlapping, pre-fetching, exploiting Shared Memory and dynamic parallelism wew employed.

### 3.1. implementation of tree

In presented implementation, which is based on C++ and CUDA the MLFMA tree is defined as an object containing a two-dimensional array. Each row of tree corresponds to a tree level and holds pointer to objects of type box. A box is representing a multi-pole. The tree object also contains four arrays, one for coordination of source and target points, and two for potential of source and target points. Potential of source points and coordintaion of both source and target points are inputs of problem, and potential of target points is unknown and output of problem. Parameter N refers to number of sourece and target points and is assumed equal for both source and target points ,i.e, problem is symmetric. Another parameter that is important in MLFMA is the clustering threshold, called CT in this article.
The tree building process starts with receiving the N, CT, input arrays (which are randomly generated), and attempting to build a valid tree in an iterative loop. In each iteration the level of the tree 'L' is increased by one starting from the default value. A tree is constructed by given L and division factor of 4. Points are then assigned to the boxes of the last level. Next it is checked wheather there is a box containing more than the CT point of the source or target points. If there was one, loop conitnues and another tree with higher levels is built. If the CT condition is satisfied, the loop is broken and the tree is used for calculations. This process is the same for both the Indexing and the Repetition methods.

### 3.2. Indexing Method

In the indexing method, each GPU thread is responsible for calculating the near field for all target points in a box. This way of distributing over threads is favorable for a single GPU device with few cores since it increases workload of each thread and thread's data might be spreaded over multiple devices. It also uses a SoA style compressed data structure in which all target points inside a box share unique neighboring source points.

In this model, data arrays are stacked in memory banks. These arrays are two arrays for sorting the coordinations of all source and target points, one array that containgin the index of the target points in each box where the corresponding indices of all the boxes are in the order of the boxe's morton index. Another array that contains the starting index of each box in the previous array, namely second order index. Another array contains the index of the source points in neighbouring of each box in consequitive morton order. Another array that contains the second order index of previous array. And finally an array contains the potential value of the source points. In reality, these seven arrays are stored in consecutive memory cells.

A GPU thread is assigned to a single box. It must first extract the second-order index of target points of its corresponding box, then extract the index of the target points in an iterative loop. In each iteration the thread extracts the second-order index of the source points corresponding to its box, and passes through them in an inner loop. Then it extracts their corrdination and their potential, and finally applies the electric potential function to a pair of target and source points. Because each thread has access to data that is located in non-consecutive memory banks (seven arrays in different memory locations), memory requests are extremely irregular. The execution speed of this method is expected to be affected by these irregular memory access patterns, subsequently leading to a high cache miss rate. In general, high cache miss rates and inefficient memory access patterns can reduce the kernel speedup to 1 or lower in compared to a single CPU.

In the data collection phase running on CPU, these seven arrays must be filled. The data collection first stores all target and source points in two arrays. Then, for each box, the index of target points and neighburing source points around it is stored. Meanwhile, the arrays responsible for the indirect indices (second-order indices) are also filled.

The execution time of this type of implementation in the CPU can be expressed by the following relation:

$$T_{IterateAndCollect_Indexing} = N(1r + 3w) + t(1r + 2w) + B(3 + N(1w) + 1w + find_nei() + 9(1r + t(1r + 1w)) + 1w) \quad (1)$$

where N is the problem size, r and w are respectively the time of a single read and write operation in RAM, find_nei() is the function of finding index of the neighbors, and B is the number of boxes. The parameter t also refers to the maximum points in the boxes after applying CT. the maximum value of t is CT and its minimum value corresponds to the case where the densest box contains CT+1 points. This box is devided to 4 smaller equal sized boxes because of CT criteria. Therefore this box would have maximum CT/4 points. The number 9 above refers to the number of adjacent neighboring boxes in the two-dimensional problem. The number of boxes is equal to $4^{L-1}$ where 4 is the branching factor of the tree and L is the height of the tree. The relation (1) is summarized as floows, assuming that the time for a single memory read and write operation is equal to m:

$$T_{iterate\ and\ collect\ indexing} = N(7m_{Indexing}) + 4^{L-1}(m_{Indexing}(11 + 19t) + 3 + find\ nei()) \quad (2)$$

The term $m_{indexing}$ above, refers to memory access time in Indexing method.
The total memory in Byte allocated for this technique is:

$$Memory_{Indexing} = 5NDouble + 4^{L-1}(2 + t + 9t)Integer = 40N + 4^{L}(2 + 10t) \quad (3)$$

Point coordination and potential values are stored as Double which is assumed to occupy 8 Bytes, and indicx values are stored as Integer which is assumed to occupy 4 Bytes. This volume of data is copied from RAM to the GPU for each execution MLFMA is executed and near-field is calculated.
The timing spent fot data transfer is important, but since it depends on the different system and runtime parameters and cannot be measured preciesly, it is refused to include it in the formulations.
The execution time of each GPU thread can also be explained using the following equation:

$$T_{Kernel_Indexing} = 4m_{Indexing} + t(3m_{Indexing} + 9t(5m_{Indexing} + O_{1-computation}) + m_{Indexing}) \quad (4)$$

where $O_{1-computation}$ is the time of the potential function that is applied between two pairs of points. In the above formula, the term t is used instead of CT because the maximum points in each box are less than or equal to t. The term t is a random variable and is not unique across the boxes, but it is certainly less than or equal to CT. The above equation simplified as follows:

$$T_{Kernel_Indexing} = 4m_{ind}(11.25t^2 + t + 1) + 9t^2 O_{1-computation} \quad (5)$$

### 3.3. Repeatition Method

In this method, each GPU thread is responsible for computing interactions between a target point and other source points in its close neighborhood. The data each thread needs to access in this method is the coordination of the target and source point, the number of source points in the nearby neighborhood, and the potential of neighboring source points. These data for all target points are collected and stacked ,i.e. stored consequitively in a single array. In this AoS way of storing data items, each thread does not need to fetch data from other parts of memory. Each target point has a maximum of nine neighboring boxes and each box contains the maximum CT source points. Therefore, the maximum number of the array elements assigned to each target point is equal to $(2 + 1 + 9 * 3 * t)$ of Double. The first two variables are the coordination of the target point, the next number is the number of nearby sources, and the next 9*3 members are 27 neighbors, including two coordinates and one potential value. All this data are stored in Double form but the second digit is read as an Integer.
With this implementation, the time to collect and transfer data to the GPU is expected to take longer than the Indexing method, and the memory access time may be longer, but in due to coalescing of memory requests, the GPU kernel should run faster and neutralize the extra time in the collection and transfer phases.
In the data collection phase in the CPU, for each box and for each target point in it, all neighboring source points are extracted and attached to an array. The execution time of the collection phase can be expressed by the following relation:

$$T_{IterateAndCollect_{Repetition}} = B(3r + t(1r + 2w + find_nei() + 1w + 9(1r + t(1r + 2w + 1r + 1w)) + 1w)) \quad (6)$$

Based on assumptions similar to those of the Indexing method, this relation can be summarized as follows:

$$T_{IterateAndCollect_{Repetition}} = 4^{L-1}(m_{Repetition} + 11tm_{Repetition} + 45m_{Repetition}t^2 + t * find_nei()) \quad (7)$$

Also, the size of the data that is generated and sent to the GPU in Bytes is equal to:

$$Memory_{Repetition} = N(3 + 27\ CT)Double = 8N(3 + 27CT) \quad (8)$$

Here the term CT is used instead of t because in GPU implementation each thread is used to calculate its starting index in a single data array without the need for additional index array. This means that the data elements are aligned to sizes 3+27*CT and there are some empty data elements.

The execution time of each GPU thread can be described as follows:

$$T_{Kernel_Repetition} = 3m_{Repetition} + 9t\left(4m_{Repetition} + O_{1-computation}\right) \quad (9)$$

Here $m_{Repetition}$ is equivalent to one memory access in the Repetition method. Here we use t instead of CT because each thread knows the number of its neighbors and can access data from mostly t neighburing source points.

### 3.4 Modeling of SpeedUp

The speed of execution caused by repetition is obtained from the following relation:

$$X_{Repetition} = \frac{T_{Indexing}}{T_{Repetition}} = \frac{T_{IterateAndCollect_Indexing} + T_{Transfer_Indexing} + T_{Kernel_Indexing}}{T_{IterateAndCollect_{Repetition}} + T_{Transfer_Repetition} + T_{Kernel_Repetition}} \quad (10)$$

where X stands for acceleration of the technique used. Since there are factors related to memory access time in the above sentences and also the execution time is also highly variable, it is difficult to calculate these expressions preciesely. However, the relation (10) can be approximated as sum of seperate speedup of collection, transfer and GPU kernel execution:

$$X_{Repetition} = \alpha X_{IterateAndCollect_Repetition} + \gamma X_{Transfer_Repetition} + \beta X_{Kerne_Reprtition} \quad (11)$$

where the values of α, β, γ are coefficients that can be evaluated in terms of execution hardware and operating system. Estimation of parameters can be achieved through applying gradient descend method to empirical data.

### 3.4.1. Modeling SpeedUp of Data Collection

Using (2) and (6) relations, data collection speedup in the replication method as follows:

$$\frac{T_{IterateAndCollect_Indexing}}{T_{IterateAndCollect_{Repetition}}} = \frac{N\left(7m_{Indexing}\right) + 4^{L-1}(m_{Indexing}(11 + 19t) + 3 + find_nei())}{4^{L-1}(3m_{Repetition} + 11tm_{Repetition} + 45m_{Repetition}t^2 + t * find_nei())} \quad (12)$$

By taking similar time for $find\ nei()$, and taking into account the larger factors, the relation (12) can be reduced to:

$$\frac{T_{IterateAndCollect_Indexing}}{T_{IterateAndCollect_{Repetition}}} \approx \frac{m_{Indexing}}{m_{Repetition}} \times \frac{1}{t} = X_m * \frac{1}{t} \quad (13)$$

where the speed of accessing RAM using the Repetition method compared to the Indexing method is denoted as $X_m$. This value is estimated as inverse ratio of the volume of the data in memory:

$$X_m \approx \lambda \frac{Memory_{Indexing} + Memory_{result}}{Memory_{Repetition} + Memory_{result}} = \lambda \frac{40N + 4^L(2 + 10t) + 8N}{8N(3 + 27CT) + 8N} \quad (14)$$

Here, λ is a coefficient that depends on the performance of RAM and shows the effect of data volume on memory access time. The value of t varies between and $\frac{CT}{4}$ and CT theoretically while the values of N and L can grow logarithmically as a function of N. For a tree of height L, there are $4^{L-1}$ boxes. Theoretically, the number of points that this tree can contain is at least $4^{L-1}$ and at most $CT * 4^{L-1}$. Taking into account the more important factors, formula (14) can be summarize as follows:

$$X_m \approx \frac{5}{8N(4+27CT)} + \frac{4^L(2+10t)}{8N(4+27CT)} + \frac{1}{4+27CT} \quad (15)$$
$$= \frac{1}{2} \times \frac{1}{D} \times \frac{10 \times \frac{1}{4}CT}{27CT}$$
$$= \frac{1}{21.6D}$$

where D is the average number of points in the box, which is obtained by dividing the total number of points by the number of boxes. By combining (15) and (13) formulation, the acceleration of data collection is expressed as follows:

$$X_{IterateAndCollect_Repetition} \approx \lambda \frac{1}{21.6tD} \quad (16)$$

According to this relation, the average density of points in the boxes, as well as the maximum points inside the boxes, have a negative effect on the efficiency of the Repetition method in the data collection phase. The lower these values, the more boxes and more evenly distributed points and the better the Repetition method will perform.

### 3.4.2 Modeling SpeedUp of the GPU kernel

The execution time of each thread have already been calculated. Because CUDA programming provides a standard interface for programming on the GPU, the programmer can define threads to more than the number of physical cores on the device. The decision of how to map these threads (virtual units) to the cores (physical units) is the responsibility of the CUDA runtime libraries. If the number of threads is greater than the number of cores, the kernel runs multiple times on device. This overhead must be calculated for both methods. To correct formulas, the calculated time must be multiplied by the number of times the kernel is executed:

$$X_{kernel_{Repetition}} = \frac{T_{Kernel_{Indexing}}}{T_{Kernel_{Repetition}}} \times \frac{Num\ Threads\ Indexing}{Total\ Device\ Cores} \times \frac{Total\ Device\ Cores}{Num\ Threads\ Repetition} \quad (18)$$
$$= \frac{T_{Kernel_{Indexing}}}{T_{Kernel_{Repetition}}} \times \frac{4^{L-1}}{N}$$

where $4^{L-1}$ depends on the number of threads in Indexing method and N depends on the number of threads in Repetition method.

By inducing (5) and (9) to (18) and using simplifications, the kernel speedup is reduced to a factor of t and the memory access speedup:

$$\frac{T_{Kernel_{Indexing}}}{T_{Kernel_{Repetition}}} = \frac{4m_{Indexing}(11.25t^2 + t + 1) + 9t^2 O_{1-computation}}{m_{Repetition}(9t+3) + 9tO_{1-computation}} \quad (19)$$

$$\leq \frac{4m_{Indexing}(t+1) + 9t^2(4m_{Indexing} + O_{1-computation})}{3m_{Repetition} + 9t(3m_{Repetition} + O_{1-computation})}$$

$$\leq \frac{4m_{Indexing}}{3m_{Repetition}} t = \frac{4}{3} X_{mem_Repetition} \times t$$

The ratio of memory access time in assumed to be equal between the Repetition and Indexing methods. Since it is difficult to accurately measure the memory access time, it is better to estimate it based on the ratio of locality of access to memory. This reclaimes that the more the data is located in close memory positions, the higher the cache hit rate, and the slower access time to memory achieves. Rather than defining a quantitative measure of locality, it is approximated by the amount of spread of memory requests:

$$X_{mem_Repetition} \approx \lambda \frac{MissRatio_{Repetition}}{MissRatio_{Indexing}} \quad (20)$$

where λ is a coefficient that depends on the GPU hardware and represents the performance of the cache and $MissRatio_{Repetition}$ is an estimation of the number of cache miss-rates or the amount of accesses to none neighburing banks of memory. The number of miss-rate of the cache for a thread is defined as follows:

$$MissRatio_{Indexing} = \frac{number\ of\ non-neighboring\ memory\ banks\ which\ a\ single\ thread\ touches}{total\ occupied\ memory\ banks} \quad (21)$$

This means that each thread experiences more letency if it accesses to more non-contigous elements of memory. Memory bank accesses in Indexing method are spreaded to non-contigous memory banks because it uses SoA data structures. In the Indexing method, each thread reads the values of 7 arrays. So each thread's read access, puts at least 7 memory banks into the cache if N and t are large enough. Therefore, an approximation of the indexing method's miss ratio is calculated based on the following equation:

$$MissRatio_{Indexing} = \frac{\left\lceil\frac{Integer}{b}\right\rceil + \left\lceil\frac{Integer}{b}\right\rceil + \left\lceil\frac{t*Integer}{b}\right\rceil + \left\lceil\frac{9t*Integer}{b}\right\rceil + \left\lceil\frac{t*2*Double}{b}\right\rceil + \left\lceil\frac{9t*3*Double}{b}\right\rceil}{\left\lceil\frac{40N+4^L(2+10t)}{b}\right\rceil} \quad (22)$$

where b is the number of bytes stored in a memory bank. In the above formula, the terms in brackets refer to reading two arrays of second ordered indices, reading the indices of the t target point and 9t source points in neighburing where all indexes are stored in Integer, reading t coordinates for target points and reading 9t coordinate and potential for neighburing source points. In the denominator, the total size of the data in memory is given according to relation (3). Assuming that for the GTX1050 device the value b is 512 Bytes, size of Double is 8 Bytes and size of an Integer is 4 Bytes, the equation (20) is simplified as follows:

$$MissRatio_{Indexing} = \frac{1+1+\left\lceil\frac{4t}{b}\right\rceil + \left\lceil\frac{36t}{b}\right\rceil + \left\lceil\frac{16t}{b}\right\rceil + \left\lceil\frac{216t}{b}\right\rceil}{\left\lceil\frac{40N+4^L(2+10t)}{b}\right\rceil} = \frac{2+\left\lceil\frac{272t}{b}\right\rceil}{\left\lceil\frac{40N+4^L(2+10t)}{b}\right\rceil} \approx \frac{272t}{40N+4^L(2+10t)} \quad (22)$$

This miss ratio is repeated for all other threads, so the value (22) must be multiplied by the number of threads in the Indexing method:

$$MissRatio_{Indexing} \approx \frac{272t}{40N+4^L(2+10t)} \times 4^{L-1} \quad (23)$$

This is an approximation of number of cache misses for Indexing method, i.e, its upper bound. For Repetition method, each GPU thread reads the same ammount of data from the entire data array, since each thread is separated from the others by adding redundancy. Therefore, the probability of a cache miss is very low and it can be roughly defined as follows:

$$MissRatio_{Repetition} = \frac{1}{N} \quad (24)$$

As a result, the ratio of cache misses in two methods is equals to:

$$\frac{MissRatio_{Repetition}}{MissRatio_{Indexing}} = \frac{1}{N} \div \left(\frac{272t \times 4^{L-1}}{40N+4^L(2+10t)}\right) \quad (25)$$

where t-factor is much smaller and is excluded from the calculations. By considering that $4^{L-1} \leq N$ the misses ratio would be:

$$\frac{MissRatio_{Repetition}}{MissRatio_{Indexing}} = \frac{40N+4^L}{N \times 4^{L-1}} \approx \frac{40}{4^{L-1}} + \frac{4}{N} \geq \frac{44}{N} \quad (26)$$

Putting relation (26) in (20) gives:

$$X_{mem_Repetition} \approx \lambda \times \frac{44}{N} \quad (27)$$

By placing (27) and (19) in equation (18), the kernel acceleration of the Repetition method is equal to:

$$X_{kernel_Repetition} = \frac{4}{3} X_{mem_Repetition} \times t \times \frac{4^{L-1}}{N} \approx \frac{4}{3}\lambda \times \frac{44}{N} \times t \times \frac{4^{L-1}}{N} \approx 55.3\lambda \times t \times \frac{1}{ND} \quad (28)$$

It can be argued that the speedup of the GPU kernel in the Repetition method decreases with increasing the size of the problem. Since the value of t is greater than D, increasing the point density in the above relation has a positive effect on the GPU kernel speed and can overcome negative effect of t.

### 3.4.3 Modeling Data Transfer SpeedUp

The data transfer time is influenced by other factors such as operating hardware and operating system, but in general it is influenced by the volume of data transferring from the RAM to the GPU. The data volume ratio is calculated in relation (15) and shows that by increasing the density of points in the boxes, the speed of the Repetition method decreases.

### 3.3.4 Analysis the Modeling of total SpeedUp

The total speedup is calculated by (11) and by placing (15) and (17) and (28) relations in that as follows:

$$X_{Repetition} = \alpha\lambda_{RAM} \frac{1}{21.6Dt} + \gamma \frac{1}{21.6D} + \beta(55.3\lambda_{GPU} \times t \times \frac{1}{ND}) \quad (29)$$

where $\lambda_{RAM}$ is related to performance of RAM and $\lambda_{GPU}$ is related to performance of GPU DRAM. The variable t plays opposite roles in accelerating the data collection and accelerating the GPU kernel. This means that Repetition method is not always faster than the Indexing method. The value $\frac{1}{D}$ can also be factored from the whole relation. It concludes that increasing the number of boxes while keeping the N constant (and therefore decreasing D), makes the Repetition method in overall faster than Indexing method. To find a balance point between the two, it is necessary to know the value $\lambda_{RAM}$ and $\lambda_{GPU}$ and coefficients α, β.

Based on the proposed analytical model, it can be suggested that in order to exploit the acceleration of Repetition method, the number of boxes must be increased while keeping N constant, and the number of sample points inside them must be reduced; or it can be said that this method is better than Indexing method for problems where the density of the sample points is low.

## 4. Evaluation of Models

In this section, validity of performance models presented in section 3 is compared with empirical data. Then advices given from that section are tested on Repetition method. The kernel function which is used here is potential function in 2D space applied on a PEC square plate which the source and target points are distributed randomly on it and the induced potential of source points into target points is calculated using MLFMA. This work focus on P2P operator which calculates near-field.

All of our tests are executed on an Intel Corei7 7thGen CPU with DDR4 16GB RAM, an NVIDIA GeForce GTX1050 and Windows10 operating system. Our code is written using CUDA 12 and C++.

### 4.1. Evaluation of Performance Models

For the accuracy of the models, the execution time of two methods is measured and compared against a reference implementation used in the work of Doriswamy[12]. In this implementation, the boxes are traversed in Morton's indexing order, and for each target point in a box, its neighboring source points are first extracted and for each source point in the $E_1$ neighborhood, the potential function is executed once and its value is added with the far-field induced potential. This implementation is chosen on the CPU as a reference, and the execution speed of the two presented techniques os compared to this baseline model.

In the baseline model, the time of data collection is the time to traverse the tree, find neighboring points and read their potential and coordination values. The computation time of the baseline model is the summation of the time that potential function applies to all pairs of source and target points. The speed of execution of methods is equal to:

$$X_{Repetition} = \frac{T_{Indexing}}{T_{Base}} \times \frac{T_{Base}}{T_{Repetition}} \quad (30)$$

Since the execution times are not precise metrics, we do not exclude it from the above relation ,i.e, the base model is executed two times because the execution time of the techniques in the same time interval is affected by the state of the operating system in that time period. Furthermore, each of GPU methods are executed once as GPU warmup process. In each experiment, we run the baseline model code, then the index model code, and then the Repetition model code so that all methods capture the same running ssystem state. For each test, the MLFMA tree is generated once and all methods are tested on the same tree.

### 4.2. Evaluation of Data Collection Performance Model

Because the data collection is done in the CPU and is very slow, the presented techniques are compared on small sized problems. Tests start with a problem of size N=5,000 to 100,000 points increasing by 5,000 at each step. Then the problem size is increased to 350,000 points with an increase of 50,000 points in each step. Default value for CT is 15 since presented technique is more efficient for sparce problems. Tree height L starts with default value of 3 and increases based on N and CT. Each test is repeated 20 times on similar tree and the running times are averaged over 20 runs.

Figure (1) shows the speedup of Repetition method based on Indexing method using formula (30) where speedup of each methods is obtained by dividing its running time by the baseline model's running time. The measured speedup is plotted with a solid black line. The vertical axis on the left shows the amount of speedup per size of the problem N. The orange dashed line is the estimated speedup using the relation (16) and the vertical axis on the right shows the corresponding speedup using relation (11). It is stated that to accurately calculate the acceleration a constant coefficient is needed. This coefficient can represent infrastructure, that is, the sensivity of RAM to irregular memory access patterns. Since two diagrams follow the same trend, the model presented in (16) can be used to predict larger problems.

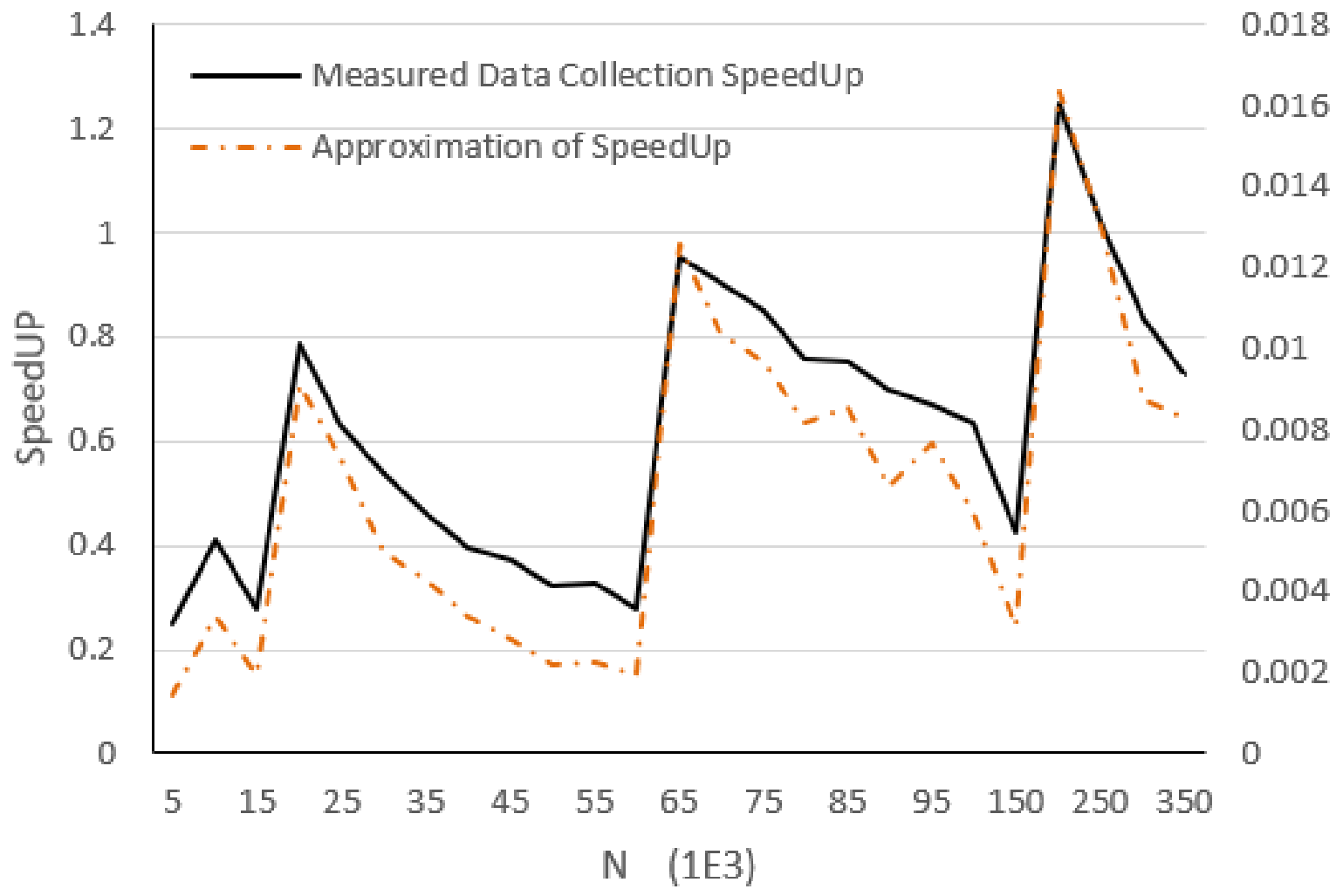

**Figure 1 – measured Repetition methods speedup (black line, left vertical axis) in data collection phase, and its estimated value using relation (16) (orange line, right vertical axis). The estimated value has similar trend with measured one.**

The value of λ for formula (16) is between 75 and 180 and the average value is 108. The reason for the oscillation of the graph (1) is related to the density. According to the relation (15), the data volume is directly related to the inverse of D. Figure (2) illusterates the average value of D in experiments. As N increases, the density and t increase moderately until $t > CT$. In this point tree level increases by one and t reduces to at least $\frac{CT}{4}$ hence density drops down to one fourth. The volume of data affected by the value of D has an important effect on the execution speed of data collection. Given this value and the relation (16), the maximum execution speedup theoretically is equal to:

$$X_{IterateAndCollec_Repetition} \approx \lambda \frac{1}{21.6tD} \geq 1 \qquad (31)$$

$$\frac{CT}{4} \sim 4 \leq t, 1 \leq D \rightarrow X_{IterateAndCollec_{Repetition}} \leq 1.25$$

According to Figure (2), the minimum values of t and D are 8 and 0.75, respectively. These values are equal to the average of 20 runs at the minimum point. According to these values, the maximum acceleration should be 1.8 times, based on empirical data:

$$8 \leq t, 0.75 \leq D \rightarrow X_{IterateAndCollec_{Repetition}} \leq 1.8 \qquad (32)$$

Figure (1) shows that by increasing the problem size to at least 200,000 or by controlling the values of D and t to small numbers, one can expect that the data collection speed of Repetition technique is at least equal to the speed of the indexing technique, then accelerating of GPU kernels would be more effective on total speedup.

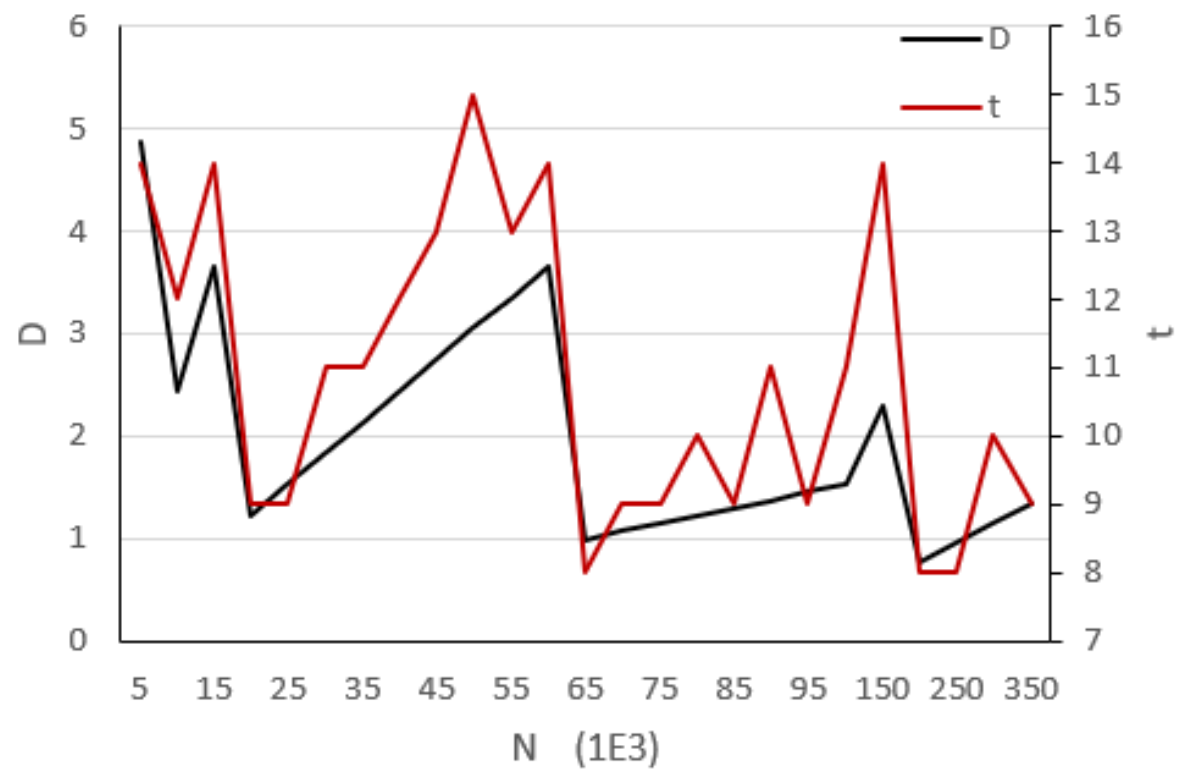

**Figure 2- Average value of D (Black line, Left Axis) and t (Red line, right axis) in experiments done for data collection speedup.**

According to Figure (2), these two variables have similar trends, and by controlling one, the other one is affected in the same way. Problems with the highest density achieves the lowest acceleration and in problems with the lowest density achieves the maximum speedup for data collection of the Repetition technique. The first case is known as low-frequency problems and the second cas problems are known as high-frequency problems.

### 4.3. Evaluation of GPU Kernel Performance Model

Since GPU kernel execution is much faster than collecting data on the CPU, in this section experiments are conducted on larger problems. In these tests N is increased by 1,000 per step beginning from 1,000 points until 100,000 points, then it increases with 50,000 points per step until 1,000,000 points. Default value for CT and L remains constant.

Figure (3) illustrates the calculated speedup of the GPU kernel of Repetition method based on Indexing method. The speedup is plotted in black solid line. The left axis shows the speedup on logarithmic scale to compensate for the effect of jumps in the value of N on the resolution of the graph. The orange dashed line shows the predicted value according to relation (28). This value has a general trend similar to the measured value, but does not follow it in partial fluctuations. One of the reasons for this error is that the runtime on the GPU does not have a strong validity.

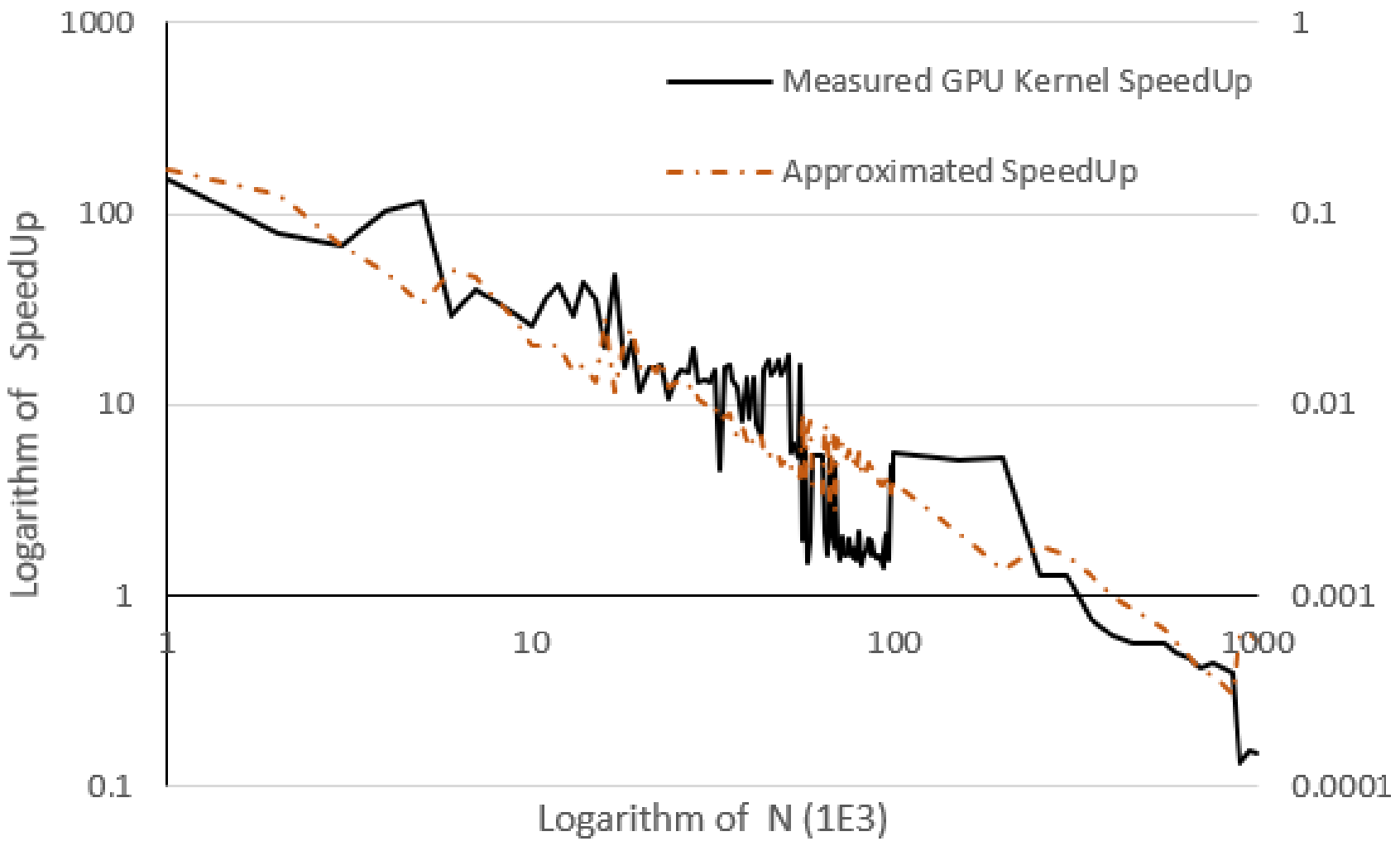

**Figure 3- measured GPU Kernel SpeedUp of Repetition method (Black Line, Left Vertical Axis) based on Indexing method, and its estimated value using relation 28(Orange Line, right vertical axis). Both axes are displayed on logarithmic scale. Trend.**

The largest value of N where Repetition method is still faster or equal to Indexing method based on formula (28) and with assuming that $t \leq CT$, is equal to:

$$X_{kernel_{Repetition}} \approx 55.3\lambda \times t \times \frac{1}{ND} \geq 1 \rightarrow N \leq 55.3\lambda \times \frac{t}{D} \quad (33)$$
$$t \leq CT = 15 \rightarrow N \leq 829.5\lambda \times \frac{1}{D}$$

The exact value of this relation requires the value of D and λ. If λ is the average value of the predicted value to the value measured in Figure (3), it is about 640. This number shows that for each unit of improvement in local memory access, the kernel runs about 640 times faster. According to figure (4) the value of D in empirical tests fluctuated between 1 and 5. The reason for the fluctuaion of D is explained earlier, in evaluation of data collection speedup performance model. Given the maximum value for D, the optimal value N (assuming that t is always 15) is:

$$106{,}000 \leq N_{Optimal} \leq 530{,}000 \quad (34)$$

According to empirical data, the optimal point is N=300,000 where t is 10 and D is 1.14 . This optimal value guides to distribute a larger size problem into smaller size of 50000~150,000 points (the half of optimal N in theoretical model and empirical data) to promote speedup of the Repetition method. In the empirical data, the speedup for N=20,000 , 50,000 and 100,000 , are 10, 3.2 and 2,5 respectively.

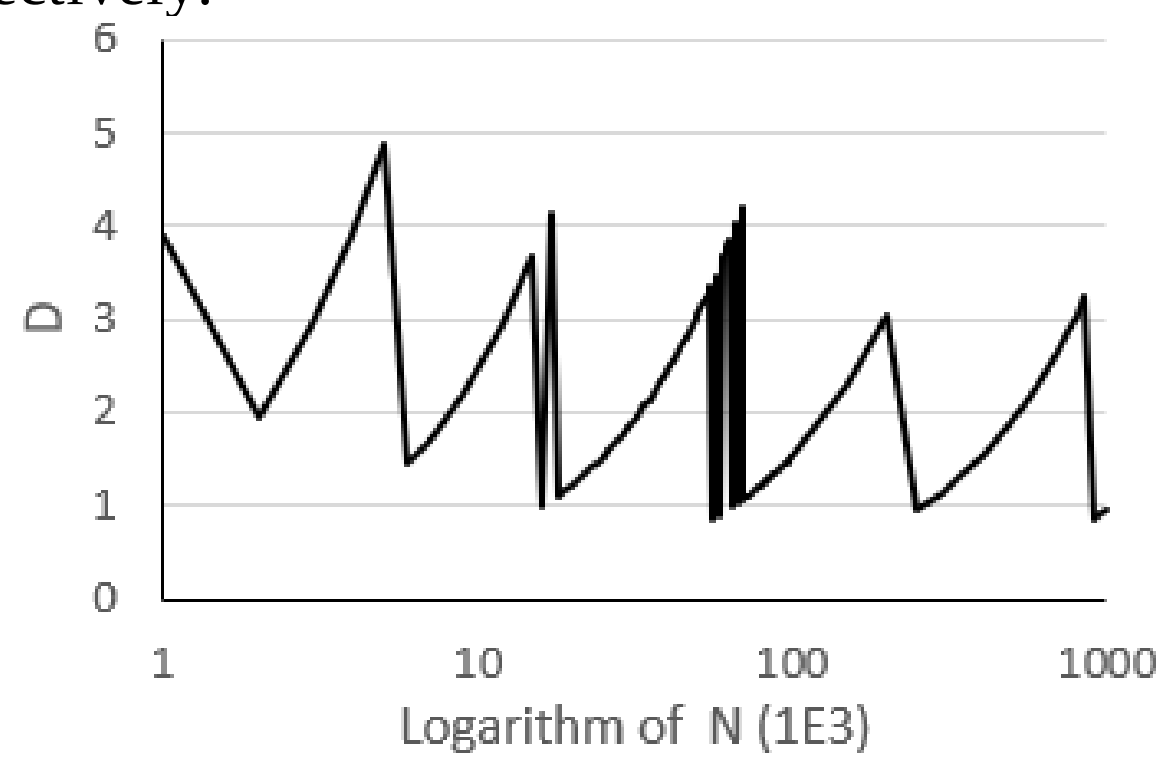

**Figure 4- Empirical value of D. Horizontal axis is displayed on logarithmic scale.**

### 4.4. Evaluation of Data Transfer Performace Model

The data transfer speed between RAM and GPU increases according to relation (15) proportional to the inverse ratio of the data volume generated during the data collection phase. But since this time is also influenced by other conditions of system setup, the relation (15) cannot be predicted well in our experiments. On the other hand, the effect of data transfer time may be hidden by the overlapping technique. Therefore, the calculation and modeling of data transfer time is not considered in this work.

### 4.5. Estimation of Modeling Coefficients

Considering the relation (29) calculation of the coefficients α, β, and γ allows to estimate the total acceleration time. These coefficients are calculated using data obtained from the test results performed for the data collection and GPU kernel execution phases. Below are the coefficients estimated using iterative methods:

$$X_{Repetition} = \alpha X_{IterateAndCollect_Repetition} + \gamma X_{Transfer_Repetition} + \beta X_{Kerne_Reprtition} \quad (36)$$
$$\alpha \approx 0.82, \beta \approx 0.09, \gamma \approx 0.18$$

In Figure (5), the total execution time (equivalent to the total time of data collection, data transfer and kernel execution) and the predicted value are shown according to coefficients (36). In this Figure the predicted values have been multiplied by empirically measuured speedups.

That is because the formulation in (29) is not accurate for transfer time. Formulation in (29) is is useful for obtaining some insights on the effect of the algorithm design and the choice of granularity on the overall speedup.

Figure (5) confirms that the idea of seperation of speedups presented in (11). The estimated coefficients have good precision according to this figure. Alignment of trends in figure (5) is very similar with the diagram (1), i.e. of data collection time is the bottleneck of the presented method. The same argument can be obtained from from coefficients in the relation (36). The GPU kernel speedup has the smallest contribution to the overall speedup. According to and figure (5) and (2), the Repetition method is slightly faster than the Indexing method when t and D are not controlled and only increased by increasing the value of N.

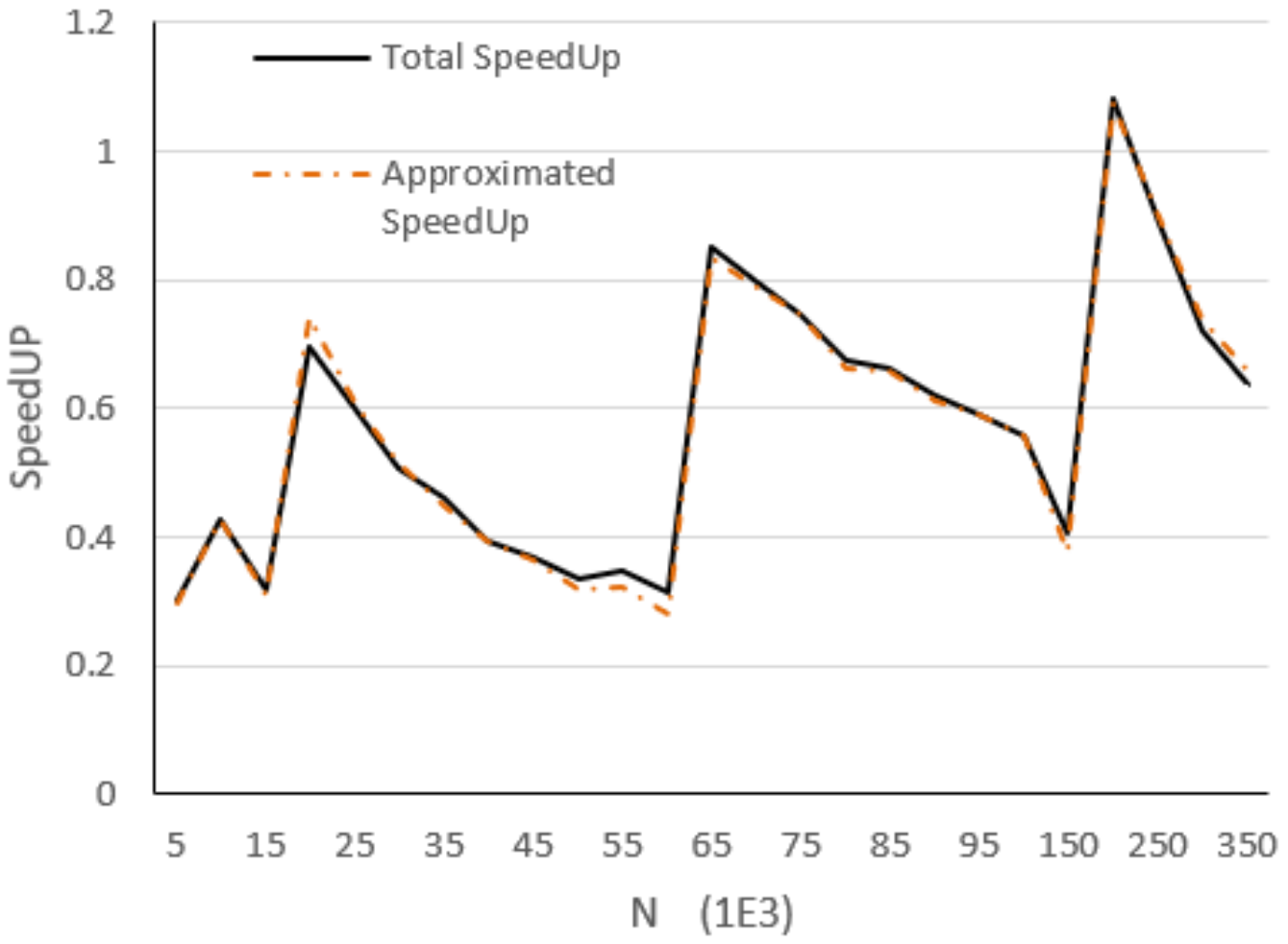

**Figure 5- total speedup of Repetition method (black line) and its estimated values according to empirical data and relation (36) (orange line).**

### 4.6. Maximizing the Total SpeedUp

Based on the results of the previous section and the relations (29) that summarize the analytical relations of this work, it is suggested to reduce the value of D and t while keeping N constant, to effectively use the repetition method in a particular problem. A simple way to do this is to increase the height of the tree by at least one unit after constructing the tree according to CT, i.e. increase the number of boxes by a factor of 4 (based on tree branching factor) and make D and t to a quarter of their original values. By increasing the height of the tree by $i$ units, according to the relation (29) and by assuming that the coefficients remain the same, the resulting acceleration will be equal to:

$$L' = L + i \rightarrow \ t' = \frac{t}{4^i}, D' = \frac{D}{4^i} \qquad (37)$$

$$X'_{Repetition} = \alpha\lambda_{RAM}\frac{4^{2i}}{21.6Dt} + \gamma\frac{4^i}{21.6D} + \beta\left(55.3\lambda_{GPU} \times t \times \frac{1}{ND}\right)$$

$$X'_{Repetition} \approx 4^{2i} \times 0.82 + 0.18 \times 4 + 0.09$$

This means that the data collection speed and thus the data transfer speedup significantly increases, while the speed of the GPU kernel remains unchanged. The formula (37) shows that by increasing one unit of tree region by one, the Repetition method becomes more than 17 times faster than the indexing method. These numbers just give us an idea, and at runtime the acceleration times are not nesserily equal to these numbers.

To evaluate the effect of varying tree height and box density, another experiment is performed this time. For Values

$$i \in \{-3,-2,-1,0,1,2,3\}$$
$$L \in \{4,5,6,7,8,9,10,11\} \quad (38)$$
different trees are generated with CT=15. The size of N in these problems is 4 times the number of boxes. The baseline model used here is the same baseline model used in the previous section.

| $L$ \ $i$ | -3 | -2 | -1 | 0 | 1 | 2 | 3 |
|---|---|---|---|---|---|---|---|
| 4 | 0.10 | 0.27 | 0.85 | 0.70 | 1.37 | 0.37 | 0.14 |
| 5 | 0.29 | 0.63 | 2.46 | 2.00 | 1.30 | 0.40 | 0.14 |
| 6 | 1.78 | 1.64 | 7.01 | 6.02 | 1.29 | 0.40 | 0.15 |
| 7 | 2.01 | 6.32 | 7.36 | 7.34 | 0.42 | 0.15 | 0.15 |
| 8 | 15.70 | 21.95 | 22.67 | 1.63 | 0.41 | 0.14 | 0.09 |
| 9 | 21.11 | 10.62 | 7.21 | 1.68 | 0.39 | 0.14 | |

Speedup of Indexing method relative to the baseline model

| $L$ \ $i$ | -3 | -2 | -1 | 0 | 1 | 2 | 3 |
|---|---|---|---|---|---|---|---|
| 4 | 1.23 | 2.16 | 2.00 | 1.60 | 1.20 | 1.53 | 1.39 |
| 5 | 2.34 | 1.99 | 1.65 | 1.77 | 1.29 | 1.16 | 1.20 |
| 6 | 2.52 | 2.58 | 2.31 | 1.73 | 1.41 | 1.39 | 1.30 |
| 7 | 1.80 | 2.38 | 1.77 | 1.93 | 1.38 | 1.21 | 1.24 |
| 8 | 2.65 | 1.95 | 2.14 | 1.56 | 1.29 | 1.24 | 1.19 |
| 9 | 1.29 | 1.84 | 1.72 | 1.44 | 1.34 | 1.02 | |

Speedup of Repetition method relative to the baseline model

| $L$ \ $i$ | -3 | -2 | -1 | 0 | 1 | 2 | 3 |
|---|---|---|---|---|---|---|---|
| 4 | 12.9 | 8.1 | 2.4 | 2.3 | 0.9 | 4.1 | 9.8 |
| 5 | 8.2 | 3.1 | 0.7 | 0.9 | 1.0 | 2.9 | 8.4 |
| 6 | 1.4 | 1.6 | 0.3 | 0.3 | 1.1 | 3.5 | 8.4 |
| 7 | 0.9 | 0.4 | 0.2 | 0.3 | 3.3 | 8.3 | 8.5 |
| 8 | 0.2 | 0.1 | 0.1 | 1.0 | 3.1 | 8.6 | 13.9 |
| 9 | 0.1 | 0.2 | 0.2 | 0.9 | 3.4 | 7.3 | |

Speedup of Repetition method compared to Indexing method

**Figure 6- Speedup of presented methods relative to baseline model, and towards each other, based on values N And L+i. Each row indicates a problem with size 4 in power of L and each column indicates the i in increasing L to L+i. The intensity of red color is corresponding to speedup of Repetition method relative to Indexing method. Points without value did not calculated due to computational resources limitations.**

In Figure (6), a comparison is made between execution time of the presented methods and baseline model, but also between them. The first sub-figure in Figure(6) shows that the Indexing method is slower than the baseline method for low-density trees where $i \geq 1$, but for high density trees in large problems where $N \geq 65{,}000$ *and* $i = -3$ it is about 22 times faster than baseline method (simple CPU implementation). The speedup of the Repetition method was always faster than the baseline model, which shows its general efficiency, but its maximum acceleration is less than 3 when $i = -3$ compared to the baseline model.
Acording to Figure (6) of the second row, the Repetition method is faster than the Indexing method in two region. The first region is when the density is high but the size of the problem is small, i.e, where $N \leq 4096$ *and* $i \leq -2$ and the problem is too small to be beneficial. According to (29) and (36), the GPU kernel which has opposite relation with N, is less affected by problem size. It can also be argued that due to the small problem and the design of the Repetition method, the data of several consecutive threads are populated in a single memory bank. This has drastically reduced the cache miss rate and increased the kernel speed. The second region is where the tree density tree decreases while the size of the problem increases ,i.e, where $N \geq 16{,}000$ *and* $i \geq 2$. In this case, the GPU kernel runs slower according to N, but the volume of data in the RAM memory decreases, and as a result, the data collection and data transfer which are mainly affected by data density and volume, become faster. This speedup is so great that the Repetition method became between 7~13 times faster than Indexing method.

## 5. Conclusion

In this work, a simple implementation of the P2P operator and a slight modification of it have been presented. Since any change in data restructuring affects the data collection time in CPU, analytical performance models have been presented to track the effect of changes in algorithm design on speedup of algorithm and find the optimal value of problem parameters. Accuracy of models were evaluated by showing their overal prediction. The optimization of MLFMA tree by controling t and D was proposed by analytical models and the result was tested empirically. According to the empirical data, the presented modification of the algorithm achives almost 13 times speedup based on unmodified algorithm for problems with more than 200,000 source and target points and 2-4 points per box.

## 6. Future Works

It is suggested to apply this modeling technique for other techniques of P2P in future works. Also it is strongly recommended to test models on various hardwares and predict device related coefficients using their hardware specifications. This bridge the gap between algorithm design and hardware measurements and helps efficiently customize algorithm for any specific hardware. Modeling of data transfer between GPU and RAM was not preciesly done in this paper, however building such precise model can complete puzzle of analytical modeling.